\documentclass[aps,prl,twocolumn,showpacs]{revtex4}
\usepackage{bm}
\usepackage{graphicx}
\begin{document}

\title{Proposed Test of Quantum Nonlocality for Continuous Variables}
\author{Hyunchul Nha and H. J. Carmichael}
\affiliation{Department of Physics, University of Auckland, Private Bag 92019,
Auckland, New Zealand} 
\date{Dec. 28, 2003}

\begin{abstract}
We propose a test of nonlocality for continuous variables using a two-mode squeezed 
state as the source of nonlocal correlations and a measurement scheme based on
conditional homodyne detection. Both the CHSH- and the CH-inequality are constructed from the conditional
homodyne data and found to be violated for a squeezing parameter larger than
$r\approx0.48$. 
\end{abstract}
\maketitle
\narrowtext

Nonlocality has been a topic of great interest ever since Bell reevaluated the claim 
of Einstein, Podolsky, and Rosen (EPR) that quantum mechanics is incomplete~\cite{Bell1}. 
Considerable effort has been invested in experimental demonstrations of nonlocality,
which manifests itself through the violation of a Bell inequality. 
Experiments for discrete variable systems have used the polarization state of photon pairs 
in an atomic cascade~\cite{Aspect}
or parametric down conversion~\cite{Zeilinger}, and the spin state of trapped ions~\cite{Rowe}.
Experiments for continuous variable (CV) systems have been rare; although they are
of particular interest as they match more closely the situation considered in the original
work of EPR. Recently, A.~Kuzmich {\it et al.}~\cite{Kuzmich} reported a signature of
nonlocality in an intensity correlation measurement of a pulsed mode EPR state. These authors
adopt the approach proposed by Grangier {\it et al.}~\cite{Grangier}, whereby an auxiliary
constraint is imposed to construct the relevant Bell inequality. Specifically, it is assumed
in this case that a change in local oscillator phase does not change the total detection
probability in the transmitted and reflected channels.

The EPR state, or in quantum optics, the two-mode squeezed state, plays a central role
in the quantum information processing of CVs~\cite{QIbook}. 
It is not possible, however, to demonstrate nonlocality for the EPR state
directly by making a CV measurement, 
since the EPR state (more generally any Gaussian state)
possesses a positive-definite Wigner function, which provides a hidden variable
model~\cite{Bell}.  For this reason, proposals for demonstrating nonlocality with 
EPR-like states (squeezed states) employ dichotomous observables derived from
some discrete physical characteristic, e.g.~even and odd photon number. In this way
it is possible to formally map the CV system onto a spin-1/2
\cite{Chen}. Alternatively, Banaszek and Wodkiewicz demonstrated that the Bell
inequality constructed from a joint parity measurement (photon number even or odd) is
violated for two-mode squeezed states~\cite{Banaszek}. They provide an operational
connection between their scheme and the Wigner phase-space distribution. These proposals
suffer from two limitations, however: first, being based on discrete data sets, they do not demonstrate  
nonlocality of a true continuous character; they are also difficult to realize experimentally,
due to the inefficiency of  photoelectric detection.
Another strategy is to retain the CV measurement in the form
of balanced homodyne detection---a highly efficient measurement compared to photon
counting---but consider a non-Gaussian state~\cite{Leonhardt}.  
Offsetting the fundamental merit of this approach is the difficulty of
preparing the states shown to exhibit nonlocality to date~\cite{Leonhardt,Gilchrist,Munro}.

In this Letter we show that the optical EPR state (two-mode squeezed state), 
although possessing a positive Wigner function, provides for a genuine CV nonlocality test 
under conditional homodyne detection, where collection of the homodyne data is conditioned
on the prior local detection of one or more photons by each of the parties sharing the EPR-correlated
fields. The detection scheme is motivated by the work of Foster {\it et al.}~\cite{Foster}.
Although the detection is conditional, the demonstrated violation of locality
may still be attributed to the EPR state itself, 
since local detection of a photon
cannot create nonlocal correlations from a classically correlated state.
 
Ralph {\it et al.}~\cite{Ralph1} also proposed a scheme demonstrating 
Bell-type correlations using homodyne detection of two-mode squeezed
states and an auxiliary measurement; but their scheme imposes a constraint on
the considered hidden-variable models~\cite{Ralph2}, in a similar manner to Refs.~\cite{Kuzmich,Grangier}.
In contrast, the underlying principle of our proposal
is that conditioning on the detection of a photon is a nonlinear operation that 
transforms a Gaussian state into a non-Gaussian one; thus, the conditioning automatically prepares a state 
that does not possess a positive-definite Wigner function. The demonstrated nonlocality may be 
attributed to the newly prepared state. Our scheme therefore constitutes a {\it genuine CV nonlocality test}, and 
demonstrates a violation of the Bell inequality in the {\it strong} sense, with 
no additional constraint on the considered hidden-variable models~\cite{Note1}. 
We show that it is insensitive to the quantum efficiency of the detectors that
generate the conditioning photocounts.

\begin{figure}
\includegraphics*[width=3.2in,keepaspectratio=true]{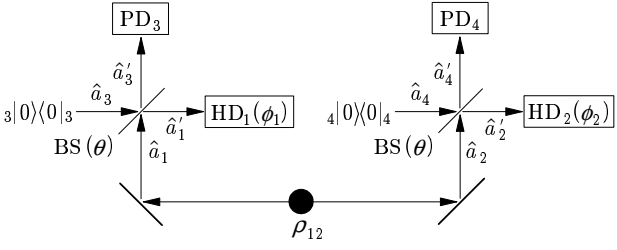}
\caption{Schematic diagram of the conditional homodyne detection of an EPR-correlated
state. Two beams of light in the correlated state $\rho_{12}$ are shared at beam
splitters ${\rm BS(\theta)}$ between photodetectors ${\rm PD}_3$ and ${\rm PD}_4$, and 
homodyne detectors ${\rm HD}_1(\phi_1)$ and ${\rm HD}_2(\phi_2)$. The homodyne data is
accepted when both photodetectors fire (each at least once).}
\label{fig:fig1}
\end{figure}

The proposed scheme is depicted in Fig.~\ref{fig:fig1}. Light in a two-mode squeezed
state $\rho_{12}$ is spatially separated and mixed with the vacuum field---states
$|0\rangle\langle0|_{3,4}$---at beam splitters of reflectance $R=\sin^2\theta$.
Balanced homodyne detection is performed at output ports $1$ and $2$,
with local oscillator phases $\phi_1$ and $\phi_2$, respectively, as shown in the
figure. Photodetectors ${\rm PD}_3$ and ${\rm PD}_4$ register photon counts, at output
ports $3$ and $4$, respectively. The photodetectors merely detect the presence
of photons, and hence output modes $3$ and $4$ are projected to states of
either zero or {\it any\/} nonzero photon number, according to the detection record.
Data from the balanced homodyne detectors, ${\rm HD}_1(\phi_1)$ and ${\rm HD}_2(\phi_2)$,
are collected only when {\it both\/} ${\rm PD}_3$ and ${\rm PD}_4$ fire. The conditional
state presented for homodyne measurement is therefore
\begin{eqnarray}
\rho_c=\frac{{\rm tr}_{34}\{\rho_{\rm BS}\hat P_3\hat P_4\}}{{\rm tr}
\{\rho_{\rm BS}\hat P_3\hat P_4\}},
\label{eqn:conditional_state}
\end{eqnarray}
with $\hat P_i=I_i-|0\rangle\langle0|_i$ ($i=3,4$) and
\begin{eqnarray}
\rho_{\rm BS}=\hat B_{13}^{\dag}\hat B_{24}^{\dag}\big(\rho_{12}\otimes|0\rangle
\langle0|_{34}\big)\hat B_{13}\hat B_{24},
\label{eqn:state} 
\end{eqnarray}
where $\hat B_{ij}^{\dag}=\exp[(\theta-\pi/2)(\hat a_i^{\dag}\hat a_j-\hat a_i
\hat a_j^{\dag})]$ accounts for the beam splitter action.

To construct a CHSH inequality we use a binning process to convert the
continuous homodyne date into binary form. We adopt the scheme used by 
Bell~\cite{Bell} and Gilchrist {\it et al.}~\cite{Gilchrist}, assigning a value
$+1$ ($-1$) when a measurement of the quadrature variable ${\hat x}_{\phi}\equiv
\frac{1}{2}\left(\hat ae^{-i\phi}+\hat a^{\dag}e^{i\phi}\right)$ gives a nonnegative
(negative) result. In the single mode case, the probability for a nonnegative result
is given by
\begin{eqnarray}
P_+(\phi)=\int_{-\infty}^\infty dx_\phi\mkern2mu p(x_\phi)\theta(x_\phi)=
\langle{\hat\theta}\left({\hat x}_{\phi}\right)\rangle,
\end{eqnarray}
with
\begin{eqnarray}
{\hat \theta}\left({\hat x}_{\phi}\right)\equiv\frac{1}{\pi}\mkern-2mu
\int_0^{\infty}\mkern-2mu dy\mkern-2mu\int_{-\infty}^{+\infty}\mkern-2mu dk\mkern2mu
e^{2ik({\hat x}_{\phi}-y)};
\end{eqnarray}
$\hat\theta(\hat x)$ is the operator step function, with $\theta(x)=1$ or
$0$, for $x\ge0$ or $x<0$. The two-mode probability, $P_{++}\left(\phi_1,\phi_2
\right)$, that measurements of ${\hat x}_{\phi_1}$ at output 1 and ${\hat x}_{\phi_2}$
at output 2 both yield nonnegative results is given by
\begin{eqnarray}
P_{++}\left(\phi_1,\phi_2\right)&=&{\rm tr}\{\rho{\hat \theta}_1\left({\hat x}_{\phi_1}
\right){\hat \theta}_2\left({\hat x}_{\phi_2}\right)\}\nonumber\\
&=&\frac{1}{\pi^2}\mkern-2mu\int\mkern-2mu d^2y\mkern-2mu\int\mkern-2mu d^2k
\mkern2mu C\left(\xi_1,\xi_2\right)e^{-2i{\bm k}\cdot{\bm y}},
\label{eqn:prob_characteristic}
\end{eqnarray}
with ${\bm y}\equiv(y_1,y_2)$, ${\bm k}\equiv(k_1,k_2)$, and $\xi_i=ik_ie^{i\phi_i}$
($i=1,2$), where
\begin{equation}
C\left(\lambda_1,\lambda_2\right)\equiv{\rm tr}\{\rho_c\hat D_1(\lambda_1)
\hat D_2(\lambda_2)\}
\label{eqn:characteristic_function}
\end{equation} 
is the characteristic function of the two-mode state $\rho_c$, with $\hat D_i(\lambda_i)
\equiv\exp(\lambda_i\hat a_i^{\dag}-\lambda^*_i\hat a_i)$ the displacement operator;
integration with respect to ${\bm y}$ and ${\bm k}$ in Eq.~(\ref{eqn:prob_characteristic})
covers the positive quadrant and the entire plane, respectively. Introducing the two-mode
Wigner function $W_{12}\left(\alpha_1,\alpha_2\right)$ [the Fourier transform of
$C(\lambda_1,\lambda_2)$] Eq.~(\ref{eqn:prob_characteristic}) yields
\begin{eqnarray}
&&P_{++}\left(\phi_1,\phi_2\right)\nonumber\\
&&=\int\mkern-2mu d^2\mkern-2mu\alpha_1\mkern-2mu\int d^2\mkern-2mu\alpha_2W_{12}
\left(\alpha_1,\alpha_2\right)F\left(\alpha_1,\phi_1\right)F\left(\alpha_2,\phi_2\right),
\mkern30mu\label{eqn:prob_Wigner}
\end{eqnarray}
where $F\left(\alpha,\phi\right)\equiv\theta\left({\rm Re}\{\alpha e^{i\phi}\}\right)$.
It is clear from this equation that if the Wigner function is positive definite, 
the amplitudes $\alpha_1,\alpha_2$ provide a hidden variable explanation 
of any correlation revealed by
$P_{++}\left(\phi_1,\phi_2\right)$ .

We find it most practical to calculate $P_{++}\left(\phi_1,\phi_2\right)$ from
Eq.~(\ref{eqn:prob_characteristic}), where we first calculate the characteristic
function (\ref{eqn:characteristic_function}). Denoting the probability
of joint photodetection by $p_{34}={{\rm tr}\{\rho_{\rm BS}\hat P_3\hat P_4\}}$ and using
the normal-ordered form $|0\rangle\langle0|_i={:\exp({-\hat a_i^{\dag}
\hat a_i}):}$, after some algebra we find
\begin{eqnarray}
C\left(\lambda_1,\lambda_2\right)&=&p_{34}^{-1}C_{\rm vac}\left(-\lambda_1\cos\theta\right)
C_{\rm vac}\left(-\lambda_2\cos\theta\right)\nonumber\\
&&\times{\rm tr}\left\{\rho'_{12}\hat D_1(\lambda_1\sin\theta) 
\hat D_2(\lambda_2\sin\theta)\right\},
\label{eqn:conditional_characteristic}
\end{eqnarray}
where $C_{\rm vac}\left(\lambda\right)=e^{-|\lambda|^2/2}$ is the characteristic function 
of the single-mode vacuum, and  
\begin{eqnarray}
\rho'_{12}=\left({\cal I}_1-{\cal S}_{1\theta}\right)\otimes\left({\cal I}_2-{\cal S}_{2\theta}
\right)\rho_{12};
\label{eqn:mapping}
\end{eqnarray}
superoperators $S_{1\theta}$ and $S_{2\theta}$ are defined by 
\begin{eqnarray}
{\cal S}_{i\theta}\equiv\sum_{k=0}^{\infty}\frac{(-\cos^2\theta)^k}{k!}\mkern2mu\hat a_i^k\cdot
\mkern2mu\hat a_i^{\dag k},
\label{eqn:superoperator}
\end{eqnarray}
while ${\cal I}_1$ and ${\cal I}_2$ are superoperator identities. The
probability $p_{34}$ is obtained by setting $\lambda_1=\lambda_2=0$ in
Eq.~(\ref{eqn:conditional_characteristic}).

The mapping $\cal M$ defined by Eqs.~(\ref{eqn:conditional_characteristic}) and (\ref{eqn:mapping}) 
has some notable features. First, it preserves local classicality, in the sense that a state 
that is locally classical in view of its Glauber P-function is mapped to another
such state. Second, and more importantly, it is a nonlinear operation
$\left({\cal M}\{\Sigma_ip_i\rho_{12}^i\}\neq \Sigma_i p_i{\cal M}\{\rho_{12}^i\}\right)$ 
that transforms a Gaussian state to a non-Gaussian one. 
In particular, a two-mode squeezed state is mapped
to a state which does not possess a positive-definite Wigner function. 
We note, for example, that the Wigner function of any single-mode field with zero
vacuum component necessarily takes negative values, since it must satisfy
$\langle0|\rho|0\rangle=2\!\int\!d^2\alpha\, W(\alpha)e^{-2|\alpha|^2}=0$.
 Thus, in spite of the
undesirable contamination by the vaccum states in Eq.~(\ref{eqn:conditional_characteristic}), 
a nonlocality test by a CV measurement becomes possible. 

The two-mode squeezed state adopted for the input is
\begin{eqnarray}
\rho_{12}=\frac{1}{\pi^2}\mkern-2mu\int\mkern-2mu d^2\lambda_1\mkern-2mu\int\mkern-2mu d^2
\lambda_2C_{\rm sq}\left(\lambda_1,\lambda_2\right)\hat D_1^{\dag}(\lambda_1)\hat D_2^{\dag}
(\lambda_2), 
\end{eqnarray}
with characteristic function
\begin{eqnarray}
C_{\rm sq}\left(\lambda_1,\lambda_2\right)&=&\exp\mkern-2mu\left[-{\textstyle\frac{1}{2}
\displaystyle}\cosh2r(|\lambda_1|^2+|\lambda_2|^2)\right.\nonumber\\
&&+\left.\vphantom{\textstyle\frac{1}{2}\displaystyle}\sinh2r{\rm Re}(\lambda_1\lambda_2^*)
\right],
\end{eqnarray}
where $e^{-2r}$ is the degree of squeezing. A long but straightforward calculation using
Eqs.~(\ref{eqn:prob_characteristic}) and (\ref{eqn:conditional_characteristic}) then gives
the binning probability  
\begin{eqnarray}
P_{++}\left(\phi_1,\phi_2\right)=\frac{1}{4}+\frac1{2\pi p_{34}}
\sum_{i=1}^3c_i\mkern-2mu\tan^{-1}\mkern-2mu\left[\frac{A}{\sqrt{b_i^2-A^2}}\right]\mkern-2mu,
\mkern10mu
\label{eqn:binning_probability}
\end{eqnarray}
with $A=\sin^2\theta\sinh2r\cos(\phi_1+\phi_2)$,
\begin{eqnarray}
b_1&=&1+2\sin^2\theta\sinh^2r,\nonumber\\
b_2&=&\sqrt{d_1d_2},\nonumber\\
b_3&=&b_1+\eta(2-\eta)\cos^4\theta\sinh^2r,
\end{eqnarray}
and 
\begin{eqnarray}
c_1&=&1\nonumber\\
c_2&=&-2(1+d_1-b_1)^{-1},\nonumber\\
c_3&=&(1+b_3-b_1+d_1-d_2)^{-1}\nonumber\\
d_1&=&b_1+\eta\cos^2\theta\sinh^2r,\nonumber\\
d_2&=&d_1-2\eta\sin^2\theta\cos^2\theta\sinh^2r.
\end{eqnarray}
These expressions take into account the nonunit quantum efficiency $\eta$ of the photodetectors
at outputs 3 and 4.  The joint photodetection probability is found
to be
\begin{equation}
p_{34}=\frac{B^2{\rm tanh}^2r\left[1+(1-B){\rm tanh}^2r\right]}
{\left[1-(1-B){\rm tanh}^2r\right]\mkern-3mu\left[1-(1-B)^2{\rm tanh}^2r\right]},
\label{eqn:joint_detection_probability}
\end{equation}
with $B=\eta\cos^2\theta$. The local probability that homodyne detection gives a nonnegative
result is $P_{i+}(\phi_i)=1/2$ ($i=1,2$), independent of the phase $\phi_i$.

We note that $P_{++}\left(\phi_1,\phi_2\right)$ depends only on the sum of local oscillator
phases $\phi_1+\phi_2$, which we denote $\phi$. Because $P_{--}\left(\phi_1,\phi_2\right)
=P_{++}\left(\phi_1,\phi_2\right)$ in the present case, the correlation
\begin{eqnarray}
E\left(\phi_1,\phi_2\right)&\equiv& P_{++}(\phi_1,\phi_2)+P_{--}(\phi_1,\phi_2)\nonumber\\
&&-P_{+-}(\phi_1,\phi_2)-P_{-+}(\phi_1,\phi_2)
\end{eqnarray}
that enters the CHSH inequality \cite{CHSH} can be evaluated from $P_{++}\left(\phi_1,\phi_2
\right)$ alone, and hence is also a function of $\phi$ only. We
write 
\begin{equation}
E(\phi)\equiv E(\phi_1,\phi_2)=4P_{++}(\phi_1,\phi_2)-1, 
\label{eqn:correlation_function}
\end{equation}
and the CHSH inequality is written as
\begin{eqnarray}
B_{\rm CHSH}\equiv|E(\phi)-E(\varphi)+E(\varphi')+E(\phi')|\le2.
\label{eqn:CHSH inequality}
\end{eqnarray}
The inequality must be satisfied for any $\phi=\phi_1+\phi_2$, $\phi'=\phi'_1+\phi'_2$,
$\varphi=\phi'_1+\phi_2$, and $\varphi'=\phi_1+\phi'_2$.  

\begin{figure}
\includegraphics*[width=3.2in,keepaspectratio=true]{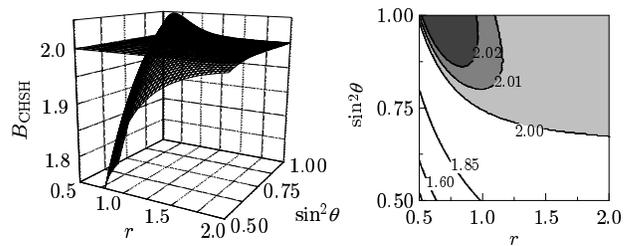}
\caption{$B_{\rm CHSH}$ as a function of the squeezing parameter $r$ and the reflectance 
$R=\sin^2\theta$, for quantum efficiency $\eta=1$. The CHSH inequality is violated for
$B_{\rm CHSH}>2$.}
\label{fig:fig2}
\end{figure}

With the phases set to $\phi=\phi'=\varphi/3=-\varphi'\equiv\Psi$, the optimal
violation occurs at $\Psi=\pi/4$. To realize this setting, we might take, for example,
$\phi_1=0$, $\phi'_1=\pi/2$, and $\phi_2=-\phi'_2=\pi/4$. Adopting these values, in
Fig.~\ref{fig:fig2} the quantity $B_{\rm CHSH}$ is plotted as a function of the
squeezing parameter $r$ and reflectance $R=\sin^2\theta$ (for $\eta=1$).
Violation of the CHSH inequality occurs for any $r$ larger than $r\approx0.48$
so long as the reflectance is sufficiently large. The need for a sufficiently large
reflectance is to be expected, since too low a reflectance contaminates the outputs
with the vacuum field from inputs 3 and 4 [Eq.~(\ref{eqn:conditional_characteristic})].
The maximal violation occurs for $r\approx0.65$, with a decreasing violation at larger
values of $r$. The appearance of such a maximum is expected also, as the mean photon number
increases with the squeezing $r$, and at large photon numbers the projections
$P_i=I_i-|0\rangle\langle0|_i$ ($i=3,4$) bring little change to the input Gaussian
state. The expected behavior is reflected in the fact that the joint photocount
probability [Eq.~(\ref{eqn:joint_detection_probability})] approaches unity for
$r\rightarrow\infty$---i.e., the conditioning is no longer selective. 

We note that whenever the CHSH inequality is violated, 
the strong CH inequality is  violated also, as was found in \cite{Munro}. 
According to the latter,
\begin{eqnarray}
B_{\rm CH}\equiv\frac{P_{++}\left(\phi\right)-P_{++}\left(\varphi\right)
+P_{++}\left(\varphi'\right)+P_{++}\left(\phi'\right)}
{P_{1+}(\phi'_1)+P_{2+}(\phi_2)}
\label{eqn:CH inequality}
\end{eqnarray}
is bounded by unity for any local realistic model \cite{Clauser}. In our case, 
$P_{1+}(\phi'_1)=P_{2+}(\phi_2)=1/2$, and hence $4B_{\rm CH}=\left(B_{\rm CHSH}+2\right)$,
from Eqs.~(\ref{eqn:correlation_function})--(\ref{eqn:CH inequality}). Thus, $B_{\rm CHSH}>2$
is equivalent to $B_{\rm CH}>1$.  

\begin{figure}
\includegraphics*[width=3.2in,keepaspectratio=true]{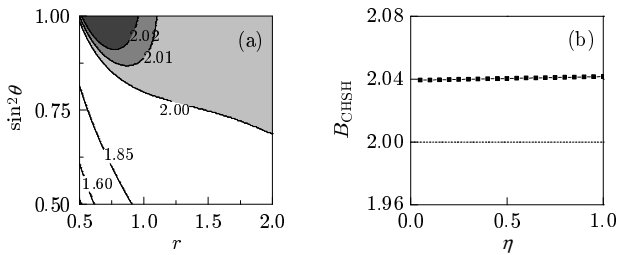}
\caption{(a) $B_{\rm CHSH}$ as a function of the squeezing parameter $r$ and the reflectance 
$R=\sin^2\theta$, for quantum efficiency $\eta=0.3$. (b) $B_{\rm CHSH}$ as a function of
$\eta$, for $r=0.6$ and $R\approx0.9891$.}
\label{fig:fig3}
\end{figure}

As discussed in \cite{Gilchrist}, homodyne detection with a strong LO is highly efficient
compared to direct photon counting. But unlike the proposal there, our
proposal relies on photodetectors ${\rm PD}_3$ and ${\rm PD}_4$ in addition to the homodyne
detectors ${\rm HD}_1$ and ${\rm HD}_2$ (Fig.~\ref{fig:fig1}). It is important, therefore, to
investigate the dependence of our results on the quantum
efficiency of these detectors. To obtain expressions
(\ref{eqn:binning_probability})--(\ref{eqn:joint_detection_probability}), a nonunit quantum
efficiency was introduced by mixing outputs 3 and 4 with additional vacuum states, placing
beam splitters of transmittance $\eta$ in front of ${\rm PD}_3$ and ${\rm PD}_4$. From
a direct calculation, the term $\cos^2\theta$ in Eq.~(\ref{eqn:superoperator})
is replaced by $\eta\cos^2\theta$, with no further change required in
Eq.~(\ref{eqn:conditional_characteristic}). Figure~\ref{fig:fig3}(a) presents a
contour plot of $B_{\rm CHSH}$ as a function of $r$ and $R=\sin^2\theta$ for a quantum
efficiency $\eta=0.3$. Comparing Fig.~\ref{fig:fig2}, we see that the violation of the
CHSH inequality is not especially sensitive to $\eta$. In particular, when $R$ is
close to unity, as in Fig.~\ref{fig:fig3}(b), $B_{\rm CHSH}$ hardly changes with $\eta$ at all;
here the CHSH inequality is violated for any quantum efficiency. Of course, a high efficiency
is desirable to maximize the rate of data acquisition, 
which is determined by the joint photocount probability $p_{34}$.

Realization of the proposed scheme appears to be feasible with currently available methods.
The EPR paradox for CVs has been demonstrated with squeezed light \cite{Ou},
and EPR-correlated light has been used for the quantum teleportation of coherent states
\cite{Furusawa}. These particular experiments involve multimode fields, however, and it
will be necessary to generalize the present treatment for application to that case.
More relevant to this work is the experiment of
Kuzmich {\it et al}.~\cite{Kuzmich}, where the optical analog of the EPR state
is produced in a {\it pulsed\/} optical parametric amplifier. In such a case the projections
associated with photodetectors ${\rm PD}_3$ and ${\rm PD}_4$ refer to the
total photon number in a single spatio-temporal mode.

Finally, we should comment on the subtle question of whether the proposed {\it conditional\/}
test demonstrates nonlocality for the initial state $\rho_{12}$. 
The most important observation in this regard is that the sampling of the homodyne data
is based on the {\it local} detection of photons coming from inputs 1 and 2. 
How observers of ${\rm PD}_3$ and ${\rm PD}_4$ determine whether to
keep or discard the data involves, at most, a classical communication. Thus, the act of
conditional data acquisition cannot create nonlocality from a classically correlated
state; this is clearly reflected in the tensor product form of Eq.~(\ref{eqn:mapping}). 
In this sense, one may argue that the demonstrated nonlocality is implicitly an attribute 
of the initial state $\rho_{12}$. In particular, the situation differs from
the so-called ``detection loophole'' \cite{Rowe}, which asserts that ``unfair'' data
sampling can lead to the violation of a Bell inequality even for a classically correlated
state. In our case the data sampling is ``fair'' and under the experimenter's control.

Of course, this is not to claim that the proposed
test demonstrates nonlocality for $\rho_{12}$ directly. Strictly, a new state having
a CV nonlocal correlation is created by the conditioning 
[Eq.~(\ref{eqn:conditional_characteristic})]
and homodyne detection is used to reveal nonlocality for the newly created state.

In conclusion, we have shown that nonlocality for CVs can be demonstrated 
by conditional homodyne detection, where the optical analog of the orginal EPR state
(two-mode squeezed state) provides the source of nonlocal correlations. Violation of the CHSH
inequality occurs for squeezing parameters greater than $r\approx0.48$, and the proposed
scheme is insensitive to the quantum efficiency of the conditioning photodetectors. 
In addition, our work has broader relevance to the field of quantum information. 
It shows that even a positive Wigner-function state can be employed 
as a source of continuous-variable nonlocality
if suitable local {\it nonlinear\/} operations are added. It provides a concrete example
of the ``photon-presence test'' performing such an operation \cite{Bartlett}. 

This work was supported by the NSF under Grant No.\ PHY-0099576 and
by the Marsden Fund of the RSNZ. Email: hnha001@postbox.auckland.ac.nz

\end{document}